\documentclass{article}
\usepackage[section]{placeins}
\usepackage{amsmath,amssymb}
\usepackage[authoryear]{natbib}
\usepackage{graphicx,color,microtype}
\usepackage{subfig}
\usepackage[pagebackref]{hyperref}

\newcommand\Rey{\mbox{\textit{Re}}}  

\title{Dynamical model of a turbulent round jet through conservation of mass flux and power}

\author
	{
	Fermin Franco$^{1, 2}$\thanks{Corresponding author: franco@sigmath.es.osaka-u.ac.jp} and Yasuhide Fukumoto$^{3}$\\
	\small{$^{1}$ Graduate School of Medicine, Osaka University}\\
	\small{2-2 Yamadaoka, Suita, Osaka, 565-0871 Japan}\\
	\small{$^{2}$ Center for Mathematical Modelling and Data Science, Osaka University}\\
	\small{1-3 Machikaneyamacho, Toyonaka, Osaka, 560-8531 Japan}\\
	\small{$^{3}$ Institute of Mathematics for Industry, Kyushu University}\\
	\small{744 Motooka, Nishi-ku, Fukuoka 819-0395, Japan}
	}
\date{September 16, 2017}

\begin{document}

\maketitle

\begin{abstract} 
We propose a family of two-phase-fluid models for a full-cone turbulent round jet that describe its dynamics in a simple but comprehensive manner with the apex angle of the cone being the main disposable parameter. The basic assumptions are that (i) the jet is statistically stationary and that (ii) it can be approximated by a mixture of two fluids with their phases in dynamic equilibrium (so-called Locally Homogeneous Flow). To derive the model, we impose either full or partial conservation of the initial mass and total power fluxes, introducing mass and power loss factors as disposable parameters. Our model equations admit implicit analytical and numerical solutions for the composite density and velocity of the two-phase fluid, both as functions of the distance from the nozzle, from which the dynamic pressure and the mass entrainment rate are calculated. Moreover, we show that the predictions of our models compare well with experimental data for single-phase turbulent air jets and atomizing liquid jets. 
\end{abstract}

\section{Introduction}\label{sec:intro}

Atomizing jets appear in a vast range of applications. A very active field of application is that of fuel jet injection engines, widely used in the automotive and aerospace industry. Other fields of application include medical apparatuses, so-called ``atomizers'' used by commercial products in many industries, flows through hoses and nozzles for various industrial purposes as well as fire fighting. Some of these applications and atomization methods are described by \citet{Jiang-etal2010}.

The earliest investigations on the hydrodynamic stability of liquid jets include the experimental works of \citet{Bidone}, who described the spatial evolution of the shape of a cross section of a liquid jet when the nozzle's orifice was not circular; and that of \citet{Savart}, who concluded two ``laws'', stating that the length of the continuous part of the jet was proportional to the jet's exit velocity, for a fixed exit diameter and, conversely, proportional to this diameter for a fixed exit velocity.

Theoretical research on the dynamics of liquid jets dates back to at least the first half of the 19th century, when \citet{Plateau} demonstrated that the surface energy of two identical liquid spheres is less than that of the equivalent volume liquid cylinder of height greater than its perimeter, which makes this cylinder unstable. However, the first one to treat this problem in a hydrodynamic stability sense was Lord \citet{Rayleigh1878} who, for an inviscid jet in vacuum, concluded that the jet is stable to non-axisymmetric disturbances but \textit{unstable} to axisymmetric disturbances of wavelength longer than the jet's perimeter. Rayleigh's theory also predicted that the most unstable wavelength, that of the greatest growth rate, is 143.7 \% the length of the jet's perimeter. 

After the pioneering theoretical work of Lord \citet{Rayleigh1878} many other early works advanced and improved upon this subject \citep{Weber1931, Tomotika1935, Taylor1962} \citep[for a review, see][]{Gorokhovski2008}. However, these results were focused on what is termed ``primary breakup'', i.e. the transition of the jet from a cylindrical geometry to the formation of the first detached droplets and ``ligaments''. This breakup has been found to depend on numerous parameters, and a characterization of the breakup mechanism, based on the dominant physical forces acting on it, has been achieved with some success. This was summarized in what is called the ``breakup regimes'' and it can be described broadly in terms of two parameters: the jet's Reynolds and the Weber numbers. Another classification which separated these breakup regimes was based on the jet's speed and the ``Z length'' or the length of the ``continuous'' part of the jet. Four main breakup regimes have been identified: (i) the capillary or Rayleigh regime; (ii) the first wind-induced regime; (iii) the second wind-induced regime; and (iv) the atomization regime. Some review articles on the topic include \citet{McCarthy-Molloy1974, Lin-Reitz1998, Liu2000, Birouk-Lekic2009} and \citet{Jiang-etal2010}.

The range of applications involving \textit{atomizing} liquid jets forming two-phase fluid flows is still large. The complexity of the atomizing process, involving numerous physical phenomena and many variables, ranging from the conditions inside the nozzle (or some generating source) to the interaction between the atomization process and the environment into which the jet is penetrating, all call for numerous challenges in physical and mathematical modelling. Notwithstanding, several mathematical models have been attempted to describe different aspects of the jets in this regime. For example, differential equations for a fuel jet's tip penetration distance as a function of time\citep{Wakuri1960,Sazhin2001,Desantes-etal05}; models for the gas entrainment rate in a full-cone spray\citep{Cossali}; and a one-dimensional model for the induced air velocity in sprays\citep{Ghosh-Hunt94}. None of these models is satisfactory by itself for understanding and application as explained below.

In this study we propose three  original related 1D mathematical models, so-called ``energy jet models'', for the macroscopic dynamics of a full-cone turbulent round jet ensuing from a circular nozzle into a stagnant fluid. This kind of jets is a key tool in many industrial processes in modern manufacturing industry \citep{Jiang-etal2010}. An advantage of our models over other analytical 1D models is that in the simplest case, the ``ideal energy jet'', it has a single experimentally measurable parameter (the half-angle $\theta$ of the jet cone) while it maintains reasonable predictive power and gives theoretical understanding that allows it to analytically calculate other physical quantities of interest. Moreover, the herein reported extended model, the ``lossy energy jet'', applies an energy conservation approach with simple turbulent and energy dissipation models, resulting in increased accuracy.
\begin{figure}
	\centering
	\includegraphics[width=0.75\columnwidth]{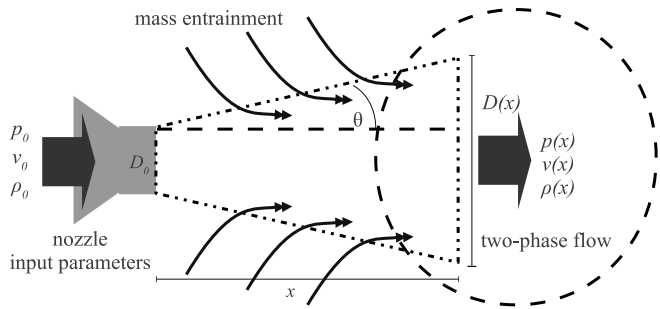}
	\caption{Diagram of the full-cone turbulent round jet geometry and the relevant physical variables.}
	\label{fig:jetdiagram}
\end{figure}

In the context of a fuel injection engine, \citet{Sazhin2001} derived a differential equation for the fuel jet's tip penetration distance as a function of time by considering a steady conical liquid fuel jet discharged into a stagnant ambient gas and imposing conservation of mass and momentum. The model relies on the restrictive assumption that the mass ratio of droplets-to-gas ($\alpha_d$, in their notation) is constant throughout the jet's length, which they then use in order to integrate the model's resulting 1st degree ODE. This assumption may be an acceptable approximation only for very small jet spread angles or for very short distances from the nozzle. Neither one of the latter requirements is met when the authors compare the equations resulting from their theory with six experimental datasets from two other sources, as well as their own \citep{Sazhin2005}, reporting jet half-angles of 13, 19, 8.5, 8.9 and 5 degrees. These angles are stated as having been calculated experimentally, the procedure not detailed or inconsistent across the different sources. Another parameter of their formula, the input velocity, is also calculated experimentally in a varied fashion across the different sources, sometimes by the mass flux rate and other times by the pressure drop at the nozzle's exit and a parameter termed the discharge coefficient, which is scattered in its recommended values across the literature, as \citet{Sazhin2001} themselves point out, ranging from 0.39 to 0.7, the latter authors having used a value of 0.8 in their only reported case. More importantly, the crucial droplets-to-gas mass ratio, $\alpha_d$, is only reported in the four experiments from the same authors \citep{Sazhin2005}, having a value of $1\times10^{-5}$. 

In the same context of a combustion engine and fuel injection, \citet{Desantes-etal05} proposed another 1D model for the diesel spray tip penetration. They focused on the dynamics of the ``main steady zone'', avoiding the transient region near the tip of the jet (the last 30$\sim$40\% of the total length) as well as the ``initial region'' near the orifice where atomization is still incomplete.  Their model relies only on conservation of momentum and on fixing the jet's geometry by means of measuring the cone angle. They make two arguable key assumptions: (i) that the behaviour of different jets with equal momentum flux and exit velocity is ``analogous''; and (ii) that the tip penetration velocity is proportional to the axial velocity in the jet's centreline. The consequence of assumption (i) is that they replace the dynamic analysis of a spray by the analysis of an assumed analogously-behaved incompressible gas jet (i.e. ignoring the droplets); they do this by exchanging the original nozzle's exit diameter for an ``equivalent diameter'' and setting the density of the jet equal to that of the gas. They also tested their model against their own experimental measurements of an air jet, but they do not state the length of the initial region near the nozzle or the position of the ``virtual origin'' of the jet (when its radius is zero), which affects the functional dependency on the axial distance variable. Also, no report is made on the initial acceleration of the fluid as we have found in \citet{FrancoMedrano-etal2017}. Thus, unlike the present study, this model has little concern in the density or liquid fraction of the spray and relies on the substitution of jet types. 
Moreover, they arrived at physical interpretations which are not necessarily derived from their model, arguing that the axial speed is twice the tip penetration because the tip loses energy by moving away stagnant air, while the fluid behind moves with the gas already in motion facing no resistance; the latter may be true but it is in no way derived from their model as it is limited to the main steady state portion of the spray. In a similar line of thought, they justified this speed using their model by assuming that the tip of the unsteady spray moves forward with a speed equal to the mean velocity of the corresponding cross-section of the steady state model, although it is to be justified that the mean cross-sectional velocity of the steady state jet corresponds to the tip penetration velocity of an unsteady jet.

\citet{Pastor-etal08} also proposed a related model for a more general case of the transient (time dependent) evolution of the tip penetration variable. They based their calculations on mass, momentum and enthalpy conservation equations. Their results cover the steady state (time independent) as a special case. However, even in this special case, their solutions are available mostly in numerical form, except in a constant-density flow case discussed in the appendix.


Another related mathematical model was made by \citet{Cossali}, where he proposed a model for the gas entrainment in a full cone spray. The main focus of the model is a non-dimensional quantity $\Lambda$ which depends on the distance $z$ from the nozzle, and expresses the ratio between gas entrainment and liquid mass flux out of the nozzle. The key ingredient of the model is a differential equation for $\Lambda$ for which Cossali found some asymptotic solutions for the near and far fields. The near-field solution was only found in explicit form as an approximation using the first term of the Taylor series centred at the position of the nozzle's exit. This approximate solution was used as a base to build another non-dimensional quantity $\Psi$ that depends linearly on the distance from the nozzle and this linear relationship was used to fit the results of eight experiments in the literature. In doing so, the value of the physical constants and even newly proposed parameters within the paper were not all used, so comparison with the experiments was limited to linear fitting the data and relating the found slope and intercept with the theoretical linear relationship from the results. The problem is that the numerous factors and constants that make up this slope and intercept contain newly proposed physical parameters that are not easy to measure experimentally (like three different mean droplet sizes, integrals over the radial distributions of gas and droplets concentrations, velocity, etc.) so this limits the predictability from the first principles.


The previous models remain untouched on a description of the density or liquid fraction of the spray, make restrictive assumptions or introduce parameters unavailable experimentally.
There are also abundant numerical models based on turbulence modelling for jets, like the one by \citet{Vallet2001} used in CFD programs like Star-CD, KIVA-3V, Ansys Fluent, as well as open source codes like OpenFOAM \citep{Stevenin2016}. 
An analytical model, producing closed expressions for experimentally measurable quantities, is advantageous to understand the parameter dependence of the phenomenon and then to predict and control it. The purpose of the present investigation is to stablish such a model from the first principles.
The models presented in this paper are an extension and generalization of the ones found in \citet{FrancoMedrano-etal2017}.

This article is organized as follows: in Sect. \ref{sec:mathmodel} we derive from physical conservation laws the basic mathematical model for the ideal energy jet. In Sect. \ref{sec:energy-loss} we introduce an energy loss factor as a simple model for enhancement of dissipation by turbulent interaction of the two phases and re-derive the state equations accordingly. In Sect. \ref{sec:mass-loss} we introduce an analogous mass loss factor accounting for mass lost across the bounded geometry of the jet cone. Next, in Sect. \ref{sec:esp-cases} we apply the state equations of the different model variations to some common limiting and special cases that recover known solutions (viz. very thin ambient fluid, (single-phase) submerged jet, and cylindrical jet without breakup). In Sect. \ref{sec:results} we compare the theoretical predictions with laboratory experiments. Finally, in Sect. \ref{sec:conclusions} we summarize the main conclusions of the present work.

\section{The ideal energy jet model}\label{sec:mathmodel}

Consider the statistically stationary state of a full-cone high-Reynolds-number turbulent jet ejecting a single-hole circular nozzle and penetrating into an ambient fluid, with constant input gauge pressure and a small conical jet spread angle. We then wish to calculate the dynamical properties of this jet at some axial distance from the nozzle. The relevant variables and parameters are depicted in Figure \ref{fig:jetdiagram}, where $\theta$ is the spread angle of the conical jet, $\rho$ the density of the fluid, $D$ the diameter of jet, $p$ the fluid's gauge pressure, $v$ the velocity of a fluid element and $x$ the axial distance from the nozzle. The subscript ``0'' indicates initial values (at the nozzle exit position), e.g. $D_0$ is the diameter of the jet at the nozzle's exit (equal to the nozzle's orifice diameter).

\subsection{Initial power}\label{sub:dotT0}

The kinetic energy of a flat disc of fluid of infinitesimal width $dx$ exiting the circular nozzle with velocity $v_0$ is
\begin{equation}\label{eq:initialenergy}
dT=\frac{1}{2}m_0v_0^2,
\end{equation}
where the mass of the flat disc is $m_0=\frac{1}{4}\rho_0\pi D_0^2dx$. Note that $dx=v_0 dt$.
This velocity may be calculated from the input gauge pressure, $p_0$, by Bernoulli's theorem, neglecting the dynamic pressure inside the nozzle and assuming that the static pressure is totally converted to the jet's dynamic pressure just outside the nozzle. Thus $p_0\approx\frac{1}{2}\rho_0v_0^2$ implies
\begin{equation}\label{eq:v0}
v_0=\sqrt{\frac{2p_0}{\rho_0}}.
\end{equation}
Note that $\rho_0$ is the density of the atomizing fluid. Substituting $m_0$ and $v_0$ from (\ref{eq:v0}) into equation (\ref{eq:initialenergy}) we get
\begin{equation}\label{eq:initialpower}
\dot{T}_0=\left.\frac{dT}{dt}\right|_{x=0}=\frac{1}{8}\pi\rho_0 D_0^2 v_0^3=\frac{1}{2}\pi D_0^2 \left(\frac{p_0^3}{2\rho_0}\right)^{\frac{1}{2}},
\end{equation}
which is the total power (energy flux per unit time) coming out of the nozzle as a result of the input gauge pressure inside the nozzle.

\subsection{Conservation of power}\label{sub:dotT}

We assume that the fluid at a distance $x$ is a two-phase fluid. The present model has been termed ``ideal energy'' to distinguish it from the ``ideal momentum'' developed by \citet{FrancoMedrano-etal2017} and the ``lossy'' models including mass and momentum or energy loss parameters to be described in Sections \ref{sec:energy-loss} and \ref{sec:mass-loss} below. Accordingly, first assume conservation of the power, i.e. the energy flux of the two-phase fluid is solely that coming from the original input pressure. The atomized fluid transfers the kinetic energy to the initially static ambient fluid by drag forces \citep{Sazhin2006, Fuchimoto-etal09} and they reach local dynamic equilibrium in such a way that both fluids move at the same speed $v$ inside the jet in a statistical sense \citep{Desantes-etal11}; this is also the main assumption under the wide class of ``Locally Homogeneous Flows'' (LHF) \citep{Faeth83, Faeth87, Faeth-etal95}. We assume that the latter process occurs so fast immediately outside the nozzle's exit that we may neglect the non-equilibrium zone near the nozzle. The latter assumption is reasonable for high-Reynolds-number pressure-atomized jets, e.g. like the ones used in real life diesel engines \citep{Sazhin2001}. This effectively allows us to treat the two-phase flow as a single fluid with a composite density $\rho$ and a single velocity $v$, depending on the distance from the nozzle, $x$.

Analogous to the calculations in subsection \ref{sub:dotT0}, the kinetic energy of a flat disc of fluid of infinitesimal width at an axial distance $x$ from the nozzle's exit is $dT=\frac{1}{2}m v^2$, where $m=\frac{1}{4}\rho\pi D^2dx$ and $dx=v dt$, and as a consequence
\begin{equation}\label{eq:impactingpower}
\dot{T}(x)=\left.\frac{dT}{dt}\right|_x = \frac{1}{8}\pi\rho D^2 v^3,
\end{equation}
which is the total power at a cross-section of the two-phase fluid jet at a distance $x$ from the nozzle.

Equating (\ref{eq:initialpower}) and (\ref{eq:impactingpower}) by conservation of the energy contained between the two planes $x=0$ and at an axial distance $x$ in steady state, and solving for $v$ we get
\begin{equation}\label{eq:vi}
v^3=\frac{1}{\rho}\left(\frac{D_0}{D}\right)^2\sqrt{\frac{8p_0^3}{\rho_0}}.
\end{equation}
This relation can be written in simpler form by using dimensionless variables as $\hat{v}^3=\hat{\rho}^{-1}\hat{D}^{-2}$, where $v$, $D$ and $\rho$ have been scaled respectively by $v_0$, $D_0$ and $\rho_0$ (the quantities at the nozzle's exit) and denoted by a hat symbol.


\subsection{Entrained volume in the two-phase fluid jet}

Assuming conservation of mass, the total fluid volume of a thin disc at the target distance is $dV=dV_0+dV_e$, where the subscript ``e'' denotes the quantities related to the ``entrained'' fluid, i.e. the total volume $dV$ of the two-phase fluid jet is just the sum of the original fluid volume coming out of the nozzle, $dV_0$, plus the added entrained fluid volume, $dV_e$, in dynamic equilibrium at distance $x$. The original volume of fluid is taken as given and, in our setting, the jet's total volume at the target distance can be calculated from the geometry. Then $dV_e=dV-dV_0$ from where we can calculate a volumetric flow rate of the mass entrainment as
\begin{equation}\label{eq:entrainment}
\frac{dV_e}{dt}=\frac{1}{4}\pi(D^2v-D_0^2v_0).
\end{equation}
Here we can substitute for $v_0$ from equation (\ref{eq:v0}) and $D=D_0+2x\tan\theta$ from the conical geometry.

\subsection{Composite density of the two-phase fluid jet}

The mean density of the two-phase fluid jet thin-disc element is just the total mass over the total volume:
\begin{equation}\label{eq:meandens}
\rho=\frac{m}{V}=\frac{dm_0+dm_e}{dV_0+dV_e}=\frac{\rho_0dV_0+\rho_edV_e}{dV_0+dV_e}
\end{equation}
After substituting $dV_e$ from (\ref{eq:entrainment}), and the initial volume element $dV_0$ from analogous calculations and solving for $\rho$, we obtain
\begin{equation}\label{eq:ri}
\rho=\rho_e+\frac{D_0^2v_0}{D^2v}(\rho_0-\rho_e),\quad\mathrm{or}\quad\hat{\rho}=\rho_*+(1-\rho_*)/\hat{D}^2\hat{v},
\end{equation}
for the density of the two-phase fluid jet at a distance $x$ from the nozzle. The second equation in (\ref{eq:ri}) is the dimensionless form for $\hat{\rho} = \rho / \rho_0$, where $\rho_*=\rho_g/\rho_0$. Fortunately, it depends on $v$ which makes the dependency circular as we can see from equation (\ref{eq:vi}). 

We implicitly assume, by calculating the \emph{mean composite density} of the two-phase fluid in Eq. (\ref{eq:meandens}), that the atomized fluid distribution throughout the disc two-phase fluid element does not differ greatly from a uniform distribution. Notice that we approximate the front of the jet by a planar front of equal density, i.e. a ``top-hat'' radial distribution. In reality this may not be the case, since the front should be spherical, and it is then in a spherical shell within the jet's cone that we should consider $\rho$ to be approximately constant, not in a plane. However, for small half-angles $\theta$ and short distances $x$ a plane should suffice for a crude estimation. The same could be said of the front's velocity $v$. Overall, we may take the above considerations as utilizing ``top-hat'' velocity and density distributions as a first order approximation. Note that slicing the spherical jet front with an $x$-normal plane provides a non-constant $\rho$ density distribution in this plane. This distribution should, however, be similar to a two-dimensional Gaussian distribution centred around the $x$-axis, i.e. the jet's ``centreline''. There are some models \citep{Ghosh-Hunt94,Desantes-etal05, Rabadi-etal07, Pastor-etal08,Desantes-etal11} which apply Gaussian velocity distributions as initial assumptions; however, this calculation will be included in a later work since we anticipate that it would not lead to a major refinement of the axial centreline quantities.

\subsection{Analytical solutions for the velocity and density}

From equations (\ref{eq:vi}) and (\ref{eq:ri}) we identify a system of two non-linear equations for two unknowns, $v$ and $\rho$. We can eliminate $\rho$ to from the system and write in dimensionless form
\begin{equation}\label{eq:nonlinear-vi}
\rho_* \hat{D}^2 \hat{v}^3+(1-\rho_*)\hat{v}^2-1=0,
\end{equation}
which is a cubic polynomial equation in $\hat{v}$, and which can be solved numerically.
%
Analogously we can get a non-linear equation for $\hat{\rho}$ eliminating $\hat{v}$ from the same described system and we obtain:
\begin{equation}\label{eq:nonlinear-ri}
\left(\frac{\hat{\rho}-\rho_*}{1-\rho_*}\right)^3=\frac{\hat{\rho}}{\hat{D}^{4}}
\end{equation}
which we can again to solve numerically.


With both $\hat{\rho}$ and $\hat{v}$ calculated, the dynamic pressure of the two-phase fluid, which accounts for the total pressure at some target axial distance $x$ from the nozzle exit, may be calculated through
\begin{equation}\label{eq:pi}
\hat{p}=\hat{\rho}\hat{v}^2,
\end{equation}
where $\hat{p}=p/p_0$, and $p_0=\rho_0 v_0^2/2$ is the initial gauge pressure as in (\ref{eq:v0}).
%

\section{Energy loss}\label{sec:energy-loss}

In this section we will generalize the above model and, instead of perfect conservation of power (energy flux), which is not realistic due to multiple energy loss processes detailed below, we assume instead that only some proportion of the energy is conserved. Consider a two-phase fluid parcel travelling from the nozzle's exit outward in an axial direction. Where physically present, we can identify the causes of energy loss of the fluid parcel as at least the following.
(i) \textit{Thermal dissipation}: some of the kinetic energy is transformed into heat due to the mutual friction between the two phases and also by viscosity within each phase. This is enhanced by turbulence.
(ii) \textit{Turbulence}: turbulent motion and energy of created vorticity in the two-phase fluid mainly due to gas entrainment at jet breakup and the turbulent wakes behind the travelling droplets.
(iii) \textit{Droplets' surface energy}: taken away at jet breakup through the droplets' surface (due to surface tension).
(iv) \textit{Escaping droplets}: kinetic energy taken away by escaping droplets expelled from the jet's bounded conical geometry by the stochastic breakup process and turbulence (notice this also produces \textit{mass loss}, although this will be addressed until section \ref{sec:mass-loss}).
(v) \textit{Droplets' oscillations}: the internal elastic motion and deformation of the droplets' surface (which may be approximated by the normal modes of oscillation of a sphere).
(vi) \textit{Droplets' rotation}: produced by some net torque at detachment during breakup.

As stated, it is clear that there are multiple sources of energy loss, but they depend mostly on the following two factors: \textit{speed} and \textit{travel time}. Alternatively, travel time at certain speed can be transformed into distance. In greater detail, the energy loss factors are affected by speed and travel time/distance in the following ways. 
(i) \textit{Thermal dissipation}: skin drag forces increase with relative velocity, so as vorticity which also produces greater dissipation through viscous forces. Dissipation also increases with travel time as we follow a two-phase fluid parcel.
(ii) \textit{Turbulence}: creation of vorticity in the two-phase fluid mainly depends on the Reynolds number, increasing with velocity. It should be fairly independent of time as the droplet formation rate.
(iii) \textit{Droplets' surface energy}: depends on the number of droplets present at some time. Droplet formation is linked to jet breakup, which for a statistically stable jet produces some approximately constant average droplet formation \textit{rate} depending on speed but independent of time since, all else remaining the same, a greater speed aids the momentum exchange between gas and liquid droplets, thus creating instability, e.g. Kelvin-Helmholtz instability \citep{Drazin}, which contributes to the breakup. Thus, the droplet formation rate and therefore the total droplet number should roughly increase with speed.
(iv) \textit{Escaping droplets}: droplets escape rate should be dependent on the magnitude of the vorticity present and also on instability growth rates which produce jet breakup; both of these factors increase with velocity, but should be again fairly independent of time.
(v) \textit{Droplets' oscillations}: this also depends on the droplets' formation rate as this energy is lost with each droplet created. Moreover, the greater the speed, the more violent the jet breakup and we should expect greater surface oscillations on the droplets. 
(vi) \textit{Droplets' rotation}: the same as for droplets' oscillations.

Therefore, it is reasonable to assume that the proportion of energy loss, due to at least the above listed factors, should depend on both the speed and travel time of each fluid element (disc). Consider two contiguous fluid elements, i.e. two discs of infinitesimal width, then the energy of the second disc, $dT_2$, should depend on the energy of the previous disc, $dT_1$, as
\begin{equation}\label{eq:energy-loss}
	dT_2=dT_1 d\mathcal{L}_e(\Delta x, v_1),
\end{equation}
where $0\leq\mathcal{L}_e\leq 1$ is the proportion of energy that is conserved after travelling a distance $\Delta x$ with speed $v_1$, where in turn $v_1$ is the velocity of the first disc. Taking the above into consideration, the overall picture is that energy loss should be greater for greater velocities and travel distances, for each fluid disc element. 

\subsection{Damping with constant velocity}

The simplest model which reproduces the above qualitative behaviour is
\begin{equation}\label{eq:energyloss-damp}
d\mathcal{L}_{e}(\Delta x, v) = \frac{1}{1+v \Delta x / H_e} = \frac{1}{1+\hat{v}\Delta\hat{x}/\hat{H_e}}, 
\end{equation}
where $H_e$ is some characteristic ``energy half-loss'' parameter at which half of the energy is lost for the given system, and $\hat{H}_e=H_e/v_0D_0=H_e/\nu_0\Rey_0$, where $\nu_0$ and $\Rey_0$ are the kinematic viscosity and Reynolds number of the liquid at the nozzle's exit, respectively. We can think of $H_e$ as the necessary distance travelled at a certain fixed speed for the fluid element to lose half its energy, in a Lagrangian point of view, or for the power through a cross section to be reduced by half, in the Eulerian frame. Notice that $H_e$ has kinematic viscosity units, $[m^2/s]$, so we may consider it to be a kinetic energy dissipation parameter, similar to a turbulent viscosity. In this sense, the model hereof derived using (\ref{eq:energyloss-damp}) may be considered to include a simple turbulence model. Notice that, the greater the velocity $v$, the greater the energy loss; this is what we call ``damping''.

\begin{figure}
	\centering
	\includegraphics[width=0.5\columnwidth]{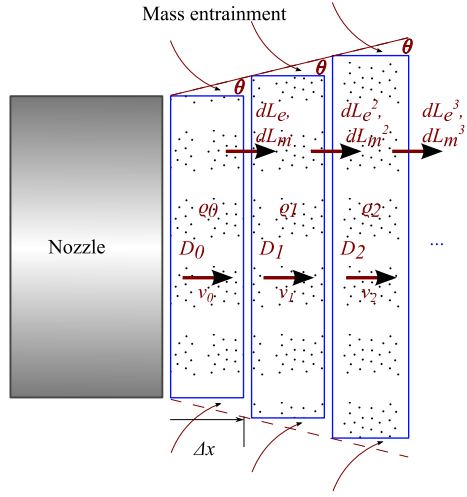}
	\caption[Slices model]{Slices model used to propagate the energy and mass loss factors in the lossy energy jet model.}
	\label{fig:slices_model}
\end{figure}

The factor $d\mathcal{L}_e$ must be considered over the disc's whole trajectory to obtain the total loss. Consider a total flight distance $x$ from the nozzle exit until the target distance. Then there will be $x/\Delta x\equiv n$ total distance intervals and their corresponding time intervals, $\Delta t$, along the jet's trajectory. If the proportion of energy preserved after one such interval is $d\mathcal{L}_e$, then $(d\mathcal{L}_e)^2$ is the proportion of energy preserved after two such intervals, \emph{considering a constant speed $v_0$}. Then the proportion of the initial energy preserved after travelling an axial distance $x$ from the nozzle exit is
\begin{equation}\label{eq:dLen}
d\mathcal{L}_e^n=\left(\frac{1}{1+H_e^{-1}v_0 x/n}\right)^n. 
\end{equation}
An sketch of this process can be seen in Figure \ref{fig:slices_model}. Proceeding to the infinitesimal limit when the disc width $\Delta x\rightarrow 0$, equivalent to $n\rightarrow\infty$, we get
\begin{equation}\label{eq:Le}
\lim_{n\rightarrow\infty} d\mathcal{L}_e^n\\
=\lim_{n\rightarrow\infty} \frac{1}{\left(1+H_e^{-1}\frac{v_0 x}{n}\right)^n}\\
=\exp(-v_0 x/H_e):=\mathcal{L}_e(x,v_0).
\end{equation}
Using this function $\mathcal{L}_e(x,v_0)$ as a proportion factor we can rewrite equations \eqref{eq:nonlinear-vi} and \eqref{eq:nonlinear-ri} by incorporating the energy loss as a function of axial distance from the nozzle exit. In dimensionless units, $\mathcal{L}_e(\hat{x})=\exp(-\hat{x}/\hat{H}_e)$.
With this modification, (\ref{eq:nonlinear-vi}) turns into
\begin{equation}\label{eq:vi-damp}
\rho_* \hat{D}^2 \hat{v}^3+(1-\rho_*)\hat{v}^2-\exp(-\hat{x}/\hat{H}_e)=0,
\end{equation}
and (\ref{eq:nonlinear-ri}) into
\begin{equation}\label{eq:ri-damp}
\left(\frac{\hat{\rho}-\rho_*}{1-\rho_*}\right)^3=\frac{\hat{\rho}}{\hat{D}^{4}}\exp(\hat{x}/\hat{H}_e)
\end{equation}
We remark that (\ref{eq:dLen}), (\ref{eq:Le}), (\ref{eq:vi-damp}) and (\ref{eq:ri-damp}) (approximately) hold only for (nearly) constant axial speed along the jet's axis. The dynamical pressure can be calculated using \eqref{eq:pi}.


\subsection{Damping with varying velocity}

The assumption of constant axial speed proved to be fairly accurate when no energy loss factor like  (\ref{eq:energyloss-damp}) is included (see Section \ref{sec:results} below). However, with inclusion of the energy loss, the velocity along the trajectory starts to decrease considerably with increasing energy loss, i.e. decreasing parameter $H_e$. This forces us to further rewrite the above equations into some recurrence formulas by which we can update the velocity value along the trajectory of the fluid element. This can be done by adapting (\ref{eq:vi}) to depend on the immediately previous value. As for the density equation, it may be kept unmodified because the amount of atomized fluid is calculated from the original state without entrained fluid and by geometrical considerations. Following a procedure analogous as in previous calculations, we obtain the velocity recurrence equation
\begin{equation}\label{eq:vi-recurrence}
\rho_* \hat{v}_i ^ 3 + \hat{D}_i ^ {-2} (1 - \rho_*) \hat{v}_i ^ 2 - \hat{v}_{i - 1} ^ 3 \left(\frac{\hat{D}_{i-1}}{\hat{D}_i}\right)^2 \frac{\hat{\rho}_{i - 1}}{1 + \hat{v}_{i - 1} \Delta \hat{x} \hat{H}_e^{-1}} = 0.
\end{equation}
Notice that \eqref{eq:vi-recurrence} reduces to \eqref{eq:nonlinear-vi} if $i=1$ and $H_e\rightarrow\infty$.

Correspondingly, the density non-linear equation \eqref{eq:nonlinear-ri} is altered to a recurrence formula
\begin{equation}\label{eq:ri-recurrence}
\left(\frac{\hat{\rho}_i - \rho_*}{1 - \rho_*}\right) ^ 3\\
=\frac{\hat{\rho}_i (1 + \hat{v}_{i - 1} \Delta \hat{x} \hat{H}_e^{-1})}{\hat{\rho}_{i - 1} \hat{v}_{i - 1}^3 \hat{D}_{i - 1} ^ 2 \hat{D}_i^4}.
\end{equation}
Notice that \eqref{eq:ri-recurrence} also reduces to \eqref{eq:nonlinear-ri} if  $i=1$ and $H_e\rightarrow\infty$.

\section{Mass loss}\label{sec:mass-loss}

We further generalize the above model and assume that only some proportion of the mass is conserved. Supposing that it exists, we can identify the causes of mass loss as at least the following. 
(i) Escaping droplets: some droplets are expelled from the jet due to the breakup and atomization processes, taking with them mass.
(ii) Evaporation: if the ambient fluid is a gas and it is not saturated in the atomized liquid's vapour, evaporation takes place and, if this vapour escapes the fixed conical geometry of the jet it reduces its total mass, albeit slowly. These both cause \emph{energy loss} as well. 
The mass loss factors are affected by velocity and travel time/distance in the following ways. (\textit{i}) Escaping droplets: droplets escape rate should be dependent on vorticity present and also on instability growth rates which produce jet breakup; both of these factors increase with velocity. (\textit{ii}) Evaporation: may occur both from the jet's continuous part (if any) and from the droplets. Evaporation turns liquid into vapour which is taken away by the surrounding gas and eventually may escape the jet. For constant temperature and gas saturation, evaporation is ``advection-driven'' and depends also on the velocity.

It is only natural then that the proportion of mass loss, due to at least the above listed factors, should depend on both the \emph{velocity} and \emph{travel time/distance} of each fluid element (disc).
Consider two contiguous fluid elements, i.e. two discs of infinitesimal width, then the mass of the second disc, $dm_2$, should depend on the mass of the previous disc as
\begin{equation}
	dm_2=dm_1 d\mathcal{L}_m(\Delta x, v_1),
\end{equation}
where $0\leq d\mathcal{L}_m\leq 1$ is the proportion of the mass that is conserved after travelling a distance $\Delta x$ at a speed $v_1$, where in turn $v_1$ is the velocity of the first disc.
Taking the above into consideration, the overall picture is that \emph{mass loss should be greater for greater velocities and greater travel distance}, for each fluid element (disc).
As in the case for the energy loss factor, the simplest model which reproduces the above qualitative behavior is
\begin{equation}\label{eq:massloss-ConstVel}
	d\mathcal{L}_m(\Delta x, v) = \frac{1}{1 + v \Delta x / H_m}=\frac{1}{1 + \hat{v} \Delta\hat{x} / \hat{H}_m}, 
\end{equation}
where $H_m$ is some \emph{characteristic ``mass half-loss parameter''} at which half of the mass is lost for the given system and $\hat{H}_m=H_m/v_0D_0=H_m/\nu_0\Rey_0$, where $\nu_0$ and $\Rey_0$ are the kinematic viscosity and Reynolds number of the liquid at the nozzle's exit, respectively. The physical meaning of $H_m$ is that it is the necessary distance travelled at a certain fixed speed for the fluid element to lose half of its mass. Notice that $H_m$ also has kinematic viscosity units, $[m^2/s]$, so we may consider it to be a mass loss parameter. Notice that, the greater the velocity, the greater the mass loss.

Considering the factor $d\mathcal{L}_m$ all along the disc's trajectory, with velocity taken as a constant, we obtain the total loss factor in an analogous way as for the energy loss:
\begin{equation}
		\lim_{n\rightarrow\infty} d\mathcal{L}_m^n =\lim_{n\rightarrow\infty} \frac{1}{\left(1 + H_m^{-1}\frac{v_0 x}{n}\right)^n}\\
		=\exp(-v_0 x/H_m):=\mathcal{L}_m(x, v_0).
\end{equation}
In dimensionless units $\mathcal{L}_m(\hat{x})=\exp(-\hat{x}/\hat{H}_m)$.

\subsection{Mass loss with constant velocity}\label{sec:ML-constvel}

Using $\mathcal{L}_m(\hat{x})$ we can rewrite our theory with the \emph{mass loss} taken into account. With this modification the velocity equation \eqref{eq:vi-damp} turns into
\begin{equation}\label{eq:vi-ML-ConstVel}
	\rho_* \hat{v} ^ 3 + \hat{D} ^ {-2} \left[(1 - \rho_*) \exp(-\hat{x}/\hat{H}_m) \hat{v} ^ 2 - \exp(-\hat{x} / \hat{H}_e)\right] = 0.
\end{equation}
Notice that this reduces to the no mass loss case \eqref{eq:vi-damp} when $H_m\rightarrow\infty$.


Likewise, the density equation \eqref{eq:ri-damp} becomes
\begin{equation}\label{eq:ri-ML-ConstVel}
\left(\frac{\hat{\rho} - \rho_*}{1 - \rho_*}\right) ^ 3 = \hat{\rho}
\hat{D} ^ {-4} \exp\left[\hat{x} \left(\frac{1}{\hat{H}_e}-\frac{3}{\hat{H}_m}\right)\right].
\end{equation}
Notice that this also reduces to the no mass loss case of \eqref{eq:ri-damp} when $H_m\rightarrow\infty$. We must remark that this calculation holds only for constant or nearly constant velocity along the jet flight.


\subsection{Mass loss with varying velocity}\label{sec:ML-varvel}

As stated before, the assumption of \textit{constant velocity} proved to be fairly accurate when \textit{no energy loss nor mass loss} factors are included. However, with the inclusion of loss factors, the velocity along the trajectory starts to vary depending on the combination of the parameters, $H_e$ and $H_m$. This forces us to further rewrite the above equations into some \textit{recurrence formulas} by which we can \textit{update the value of the velocity} along the trajectory of the fluid element. This is done by adapting the equations to depend on an immediately previous value.
Following an analogous procedure as in previous calculations, we obtain the velocity recurrence equation
\begin{equation}\label{eq:vi-ML-recurrence}
	\rho_* \hat{v}_i ^ 3 + \left(\frac{\hat{D}_{i-1}}{\hat{D}_i}\right) ^ 2 \hat{v}_{i - 1} \left(\frac{\hat{\rho}_{i - 1}}{1 + \hat{v}_{i - 1} \Delta \hat{x} \hat{H}_m ^ {-1}} - \rho_* \right) \hat{v}_i ^ 2 - \hat{v}_{i - 1} ^ 3 \left(\frac{\hat{D}_{i - 1}}{\hat{D}_i}\right) ^ 2 \frac{\hat{\rho}_{i - 1}}{1 + \hat{v}_{i - 1}\Delta \hat{x} \hat{H}_e^{-1}} = 0.
\end{equation}
Notice that \eqref{eq:vi-ML-recurrence} also reduces to \eqref{eq:nonlinear-vi} without loss factors if $H_m,\,H_e\rightarrow\infty$ and $i=1$. As for the density, we find that
\begin{equation}\label{eq:ri-ML-recurrence}
\left(\frac{\hat{\rho}_i - \rho_*}{\frac{\hat{\rho}_{i - 1}}{1 + \hat{v}_{i - 1}\Delta \hat{x} \hat{H}_m ^ {-1}} - \rho_*}\right) ^ 3 = \\
\left(\frac{\hat{D}_{i - 1}}{\hat{D}_i} \right) ^ 4 \frac{\hat{\rho}_i \left(1 + \hat{v}_{i-1} \Delta \hat{x} \hat{H}_e^{-1}\right)}{\hat{\rho}_{i-1}}.
\end{equation}
Notice that \eqref{eq:ri-ML-recurrence} also reduces to \eqref{eq:nonlinear-ri} without any loss if  $i=1$ and $H_m,\,H_e\rightarrow\infty$.

\subsection{Liquid-only mass loss}\label{sub:LOML}

In the previous modelling of the mass loss factor we implicitly considered mass loss to happen from both phases of the jet. In the case of an atomized liquid and an ambient gas, the latter is \emph{not} realistic, since we expect that the mass loss mechanisms affect only the liquid phase, as was described in the enlisted physical mechanisms above. An ambient gas does not lose any mass. Therefore, in this case we may restrict the mass-loss to occur only for the liquid phase and get a different set of equations.

First notice that the mass loss assumption implies a volume loss relationship
\begin{equation}\label{eq:volume-loss}
	V_j^l = V_{j-1}^l d\mathcal{L}_m(\Delta x, v_{j - 1}),
\end{equation}
so that the volume of entrained ambient gas at any step can be calculated as
\begin{equation}\label{eq:gas-volume}
	V_j^g  	= V_j - V_j^l\\
	= V_j - V_{j-1}^l d\mathcal{L}_m
\end{equation}
where the superscripts ``g'' and ``l'' refer to the quantities for gas and liquid, respectively; and, if no superscript is written, the quantity refers to the composite two-phase fluid. The subscript ``$j$'' indicates the ``$j$-th step'' position. Notice that $V_j^l$ for $j>0$ cannot be calculated as the volume of a cylinder as with $V_0^l$ or $V_j$, since the shape of the distributed droplets inside the jet is \textit{irregular}.
However, we may calculate $V_j^l$ recursively from \eqref{eq:volume-loss}, since we know the total volume, $V_j=\frac{1}{4}\pi D^2 \Delta x$, we also know the initial liquid volume $V_0^l=\frac{1}{4}\pi D_0^2 \Delta x$.
Separating the mass loss to affect only the liquid phase, we may calculate again the mean composite density as
\begin{equation}\label{eq:LOML-compdens}
	\rho_j 	= \frac{\rho_0 V_j^l + \rho_a V_j^g}{V_j}.
\end{equation}
If we substitute \eqref{eq:gas-volume} into \eqref{eq:LOML-compdens} then simplifying we have
\begin{equation}\label{eq:density}
	\rho_j = \rho_g + \frac{V_{j}^l}{V_j}(\rho_0 - \rho_g).
\end{equation}
Notice that $\rho_j$ can be calculated directly from this formula by first calculating the values of $V_j$ from the geometry and $V_j^l$ from \eqref{eq:volume-loss}.

As in the case where we introduced the energy loss (damping) with varying velocity in Section \ref{sec:energy-loss}, we should have the same equation for the velocity:
\begin{equation}\label{eq:velocity}
	v_j^3 = \left(\frac{D_{j-1}}{D_j}\right)^2 \frac{\rho_{j-1}}{\rho_j}
	v_{j-1}^3 d\mathcal{L}_e(\Delta x, v_{j - 1}).
\end{equation}
Eliminating $\rho_j$ from \eqref{eq:density} and \eqref{eq:velocity}:
\begin{equation}\label{eq:vel-solved}
	v_j = \left(\frac{D_{j - 1}}{D_j}\right) ^ {2 / 3}
	(\rho_{j - 1} d\mathcal{L}_e) ^ {1 / 3} \rho_j v_{j - 1}.
\end{equation}
From \eqref{eq:vel-solved} we can calculate $v_j$ directly having previously calculated $\rho_{j - 1}$ and $\rho_j$ from \eqref{eq:density}.

Notice that the velocity recurrence relation \eqref{eq:vel-solved} can be written as
\begin{equation}
	v_j^3 	= \frac{\frac{1}{4} \pi D_{j - 1} ^ 2 v_{j - 1} dt \rho_{j - 1}}
	{\frac{1}{4} \pi D_{j} ^ 2 v_j dt \rho_{j}} v_{j - 1} ^ 2 d\mathcal{L}_e
	= \frac{V_{j - 1} \rho_{j - 1}}{V_j \rho_j} v_{j - 1} ^ 2 d\mathcal{L}_e.
\end{equation}
And then, rearranging terms, we obtain
\begin{equation}
	\frac{1}{2}V_j\rho_jv_j^2 = 
	\frac{1}{2}V_{j-1}\rho_{j-1}v_{j-1}^2 d\mathcal{L}_e,
\end{equation}
i.e. $T_j = T_{j-1} d\mathcal{L}_e$, which just states the conservation of a proportion of the kinetic energy as originally proposed in \eqref{eq:energy-loss}.


\section{Limit and special cases}\label{sec:esp-cases}

In this section we present some special or limiting cases of the above proposed mathematical models. The results of Section \ref{sec:ML-constvel} generalize all the analytical solutions of the previous sections, and therefore it suffices to analyse the equations of the cases therein. It is worth remembering for the following analysis that $\hat{D}=1 + 2 \hat{x} \tan\theta$ is the diameter of the jet.

\subsection{An atomizing jet in a very thin atmosphere}

With this assumption, $\rho_*\approx 0$, which physically means a jet inside a very thin atmosphere.
We see from Equation (\ref{eq:vi-ML-ConstVel}), after simplification, that the state equation for the velocity may be written
\begin{equation}\label{eq:vi-vacuum}
\hat{v} = \exp\left[-\frac{\hat{x}}{2}\left(\frac{1}{\hat{H}_e}-\frac{1}{\hat{H}_m}\right)\right],
\end{equation}
which gives $\hat{v}=1$ for $\hat{H}_{e, m}\rightarrow\infty$, i.e. the velocity remains constant at the same initial value, being consistent with a physical intuition that the liquid droplets do not lose any energy to the surrounding thin gas in case nearly no drag forces nor turbulent two-phase flow present. Note that the velocity is maintained for $H_e=H_m$. In a Lagrangian point of view, if the proportion of energy lost by a fluid parcel is the same as that of its lost mass, the mass remaining within the parcel carries the same energy per unit mass, i.e. squared velocity, independently of the axial position and thus the velocity is everywhere constant. On the other hand, if $H_m > H_e$, mass loss is proportionally \textit{smaller} than energy loss (remember that for constant velocity $H_{m,e}$ can be interpreted as half-loss \textit{distances}), and the velocity decays with the axial distance and tends to zero at infinity, which admits and interpretation that can be interpreted as proportionally more power is lost than mass flux, causing the power per unit mass to decrease within the jet. The case $H_e<H_m$ is physically impossible since it implies that the velocity increases with axial distance and tend to infinity, i.e. the jet accelerates. Therefore, this case helps us establish that $H_m\geq H_e>0$. 
From Equation (\ref{eq:ri-ML-ConstVel}) we get the state equation for the density,
\begin{equation}\label{eq:ri-vacuum}
\hat{\rho} = \frac{1}{\hat{D}^{2}}\exp\left[\frac{\hat{x}}{2}\left(\frac{1}{\hat{H}_e}-\frac{3}{\hat{H}_m}\right)\right],
\end{equation}
where we can see that for $\hat{H}_{e, m}\rightarrow\infty$ we have $\hat{\rho}=\hat{D}^{-2}$ and $\rho\rightarrow0$ as $x\rightarrow\infty$ in this case. Interestingly, $H_m=3 H_e$ gives the same result. In any case, $\hat{\rho}=1$ when $\hat{x}=0$, in accordance with the fact that at the nozzle we have only liquid.
For the solution (\ref{eq:ri-vacuum}) to be physically acceptable, we should impose that $\hat{\rho}'(\hat{x})\leq0$, i.e. the composite density of the two phase fluid should be a non-increasing function of the axial distance $x$. Let $\kappa=(\hat{H}_e^{-1} - 3\hat{H}_m^{-1})/2$. Then $\hat{\rho}=\hat{D}^{-2}\exp\kappa\hat{x}$ and since $\exp\kappa\hat{x}>0$, we can prove that if
\begin{equation*}
\kappa < \frac{4\tan\theta}{1 + 2a\tan\theta},
\end{equation*}
then $\hat{\rho}'<0$ for all $\hat{x}<a\in\Re^+$. In particular, if $\kappa<0$ then $\hat{\rho}'<0$ for all $\hat{x}>0$. Therefore, we conclude that in order to have a physically acceptable solution, the loss parameters should verify
\begin{equation}
0<\hat{H}_e\leq\hat{H}_m\leq 3\hat{H}_e.
\end{equation}
Since in this special case both $v$ and $\rho$ are given as an explicit analytic solution in \eqref{eq:vi-vacuum} and \eqref{eq:ri-vacuum}, respectively, we can also calculate such a solution for the pressure from \eqref{eq:pi} as
\begin{equation}\label{eq:pi-vacuum}
\hat{p}=\hat{D}^{-2}\exp\left[-\frac{1}{2}\hat{x}\left(\hat{H}_e + \hat{H}_m\right)\right],
\end{equation}
indicating that $p=p_0$ at the nozzle, and $p\rightarrow 0$ as $x\rightarrow +\infty$. Notably, any combination of the loss factors necessarily induces a decay in the dynamic pressure with axial distance and the only way to neutralize the loss factor is by taking both $\hat{H}_{e, m}\rightarrow\infty$, which is the ideal case $\hat{p}=\hat{D}^{-2}$.


\subsection{The submerged jet}

We look into a single-phase fluid jet, i.e. liquid or gas only, by taking $\rho_*=1$. In view of \eqref{eq:ri-ML-ConstVel} we have $\hat{\rho}=1$, as expected; the density is just that of the common fluid.
From \eqref{eq:vi-ML-ConstVel} we get the state equation for the velocity
\begin{equation}\label{eq:v-subjet}
\hat{v}=\hat{D}^{-2/3}\exp\left(-\hat{x}/\hat{H}_e\right),
\end{equation}
which gives $v=v_0$ for $x=0$, i.e. at the nozzle exit, and $v\rightarrow 0$ as $x\rightarrow +\infty$, as it should be. Notice that in this case the density remains constant and velocity decays with distance whilst for the case of a thin atmosphere we had the converse, viz. the velocity remains constant and the density decays.
From (\ref{eq:pi}) combined with (\ref{eq:v-subjet}) we obtain the state equation for the pressure
\begin{equation}\label{eq:p-subjet}
\hat{p}=\hat{D}^{-4/3}\exp\left(-2\hat{x}/\hat{H}_e\right),
\end{equation}
In both \eqref{eq:v-subjet} and \eqref{eq:p-subjet} the mass loss factor disappears, which agrees with the physical interpretation that there cannot be any mass loss in a single phase fluid jet, as the both mechanisms described in section \ref{sec:mass-loss} are non existent, viz. escaping droplets and evaporation. On the other hand, energy loss is retained but if $\hat{H}_e\rightarrow\infty$ then $\hat{p}=\hat{v}^2=\hat{D}^{-4/3}$.


\section{Comparison with experiments and discussion}\label{sec:results}

\begin{table}
	\centering
	\begin{tabular}{ccc}
		\hline
		$\rho_a = 1.1644$ (kg/m$^3$) & $\theta$ = 1/18 ($^\circ)$ &  $p_0 = 5, 10, 15$ (MPa)\\ \hline
		$\rho_0 = 995.6502$ kg/m$^3$ & $D_0$=1.6 (mm) & $T$ = 30 ($^\circ$ C) \\ \hline
		$x$ = 0.3 (m) & & \\ \hline
	\end{tabular}
	\caption[Physical parameters used in the computations.]{Physical parameters used in the computations for a water jet in air using the lossy energy models.}
	\label{tab:parameters}
\end{table}

\begin{figure}
	\centering
	\includegraphics[width=1.1\columnwidth]{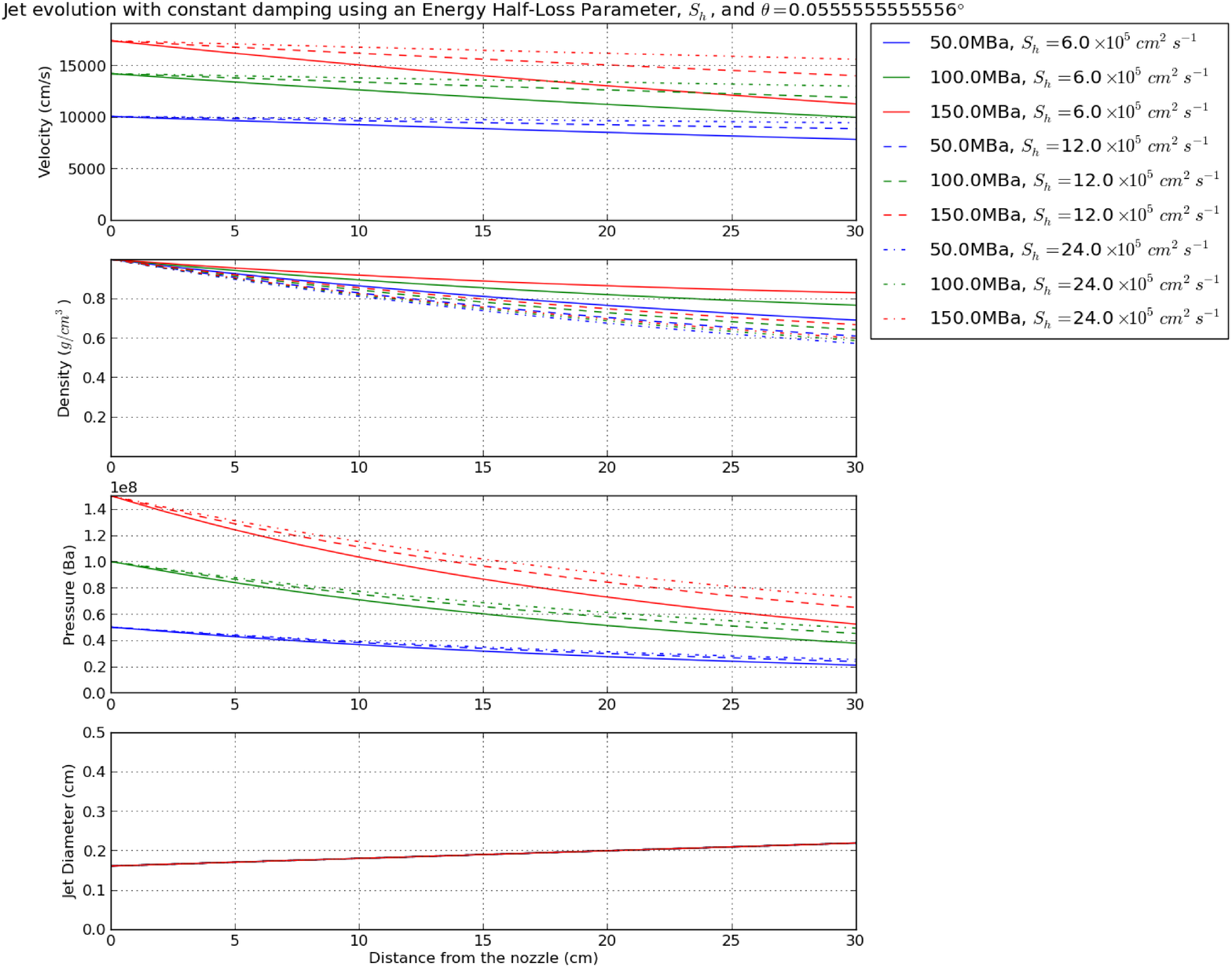}
	\caption{Results for energy loss with damping at constant velocity.}
	\label{fig:constantdamping}
\end{figure}

\begin{figure}
	\centering
	\includegraphics[width=1.1\columnwidth]{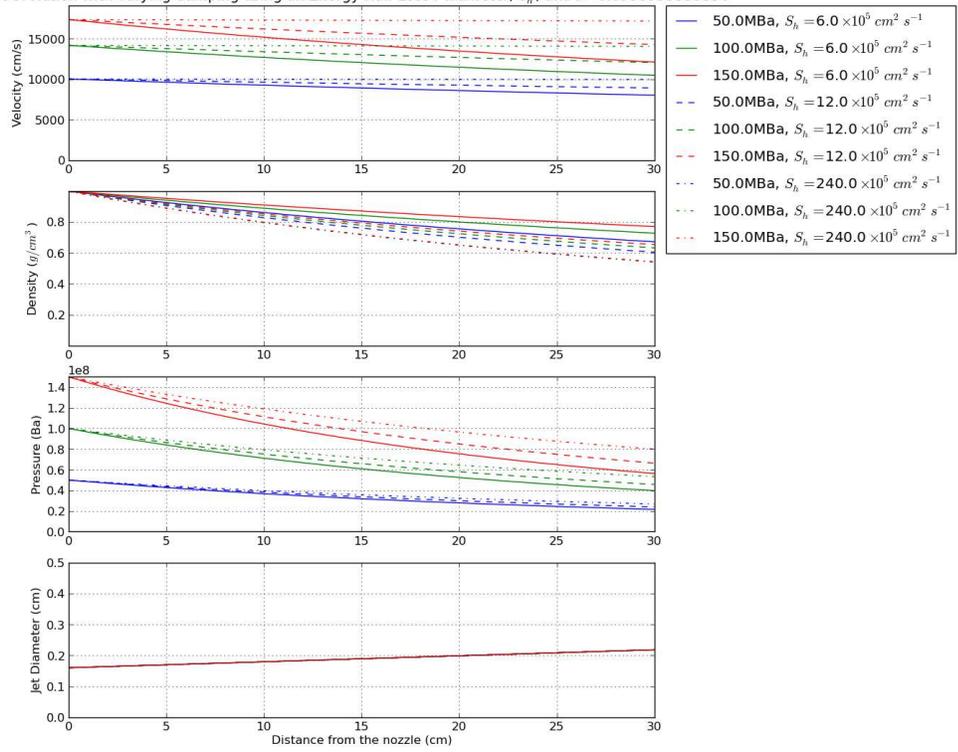}
	\caption{Results for energy loss with damping at varying velocity.}
	\label{fig:varyingdamping}
\end{figure}

Using the physical parameters in table \ref{tab:parameters} we shall solve for the velocity, density, dynamic pressure and diameter of a conical high-speed turbulent atomizing liquid jet, which bears close resemblance with laboratory parameters. The input pressure $p_0$ is chosen to be the same as used in the laboratory experiments, as are the nozzle diameter and the jet flight distance before impact. The temperature is chosen as an ambient temperature of 30 $^\circ$C. A special care should be exercised for the jet's cone spread angle. In laboratory measurements this angle is about one-sixth (1/6) of a degree. However, the measured angle corresponds vaguely to the spread angle of the jet as a whole, consisting of both the core jet and a ``halo'' of droplets. The angle required for the theoretical calculations in this study is the angle of the \textit{core} of the jet and not the whole jet.  After a trial and error, the desired angle is found to be approximately one-third of the reported experimental halo; i.e. one-eighteenth (1/18) of a degree and fits the experimental data most closely with constant speed. The result of the theoretical calculation for the energy loss factor, damping with constant speed, is shown in figure \ref{fig:constantdamping}. The result of the theoretical calculation for the energy loss factor and damping with varying velocity are shown in figure \ref{fig:varyingdamping}. The result of the theoretical calculation for the energy loss factor, damping with varying velocity and liquid-only mass-loss, is shown in figure \ref{fig:liquidonly}. Remarkably, comparison of the most accurate results of figure \ref{fig:varyingdamping} with the experimental data shows an almost perfect fit for the dynamic pressure of the jet, with the choice of the energy half-loss parameter of $H_e=12\times 10^5\, cm^2/s$ (in the figure, $\gamma=H_e$).

%

\begin{figure}[tbph]
	\centering
	\includegraphics[width=0.7\linewidth]{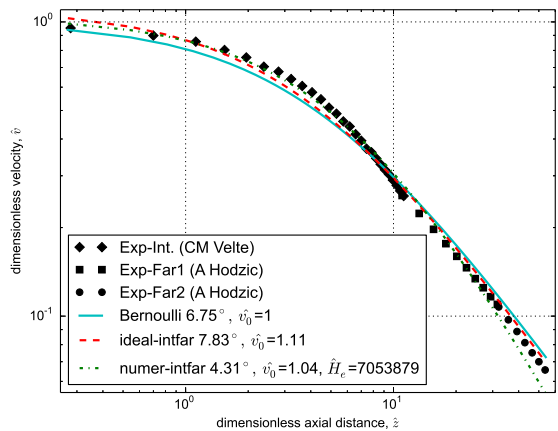}
	\caption[Centreline velocity fits]{Centreline velocity fits to experimental data for a single-phase air jet as reported in \citet{FrancoMedrano-etal2017}.}
	\label{fig:fits_intfar_loglog-square}
\end{figure}

Figure \ref{fig:fits_intfar_loglog-square} shows the centreline velocity fits to experimental data for a single-phase air jet as reported in \citet{FrancoMedrano-etal2017}. Each of the three curves represents the theoretical results of a different model. Both the continuous line and the dashed line belong to the ideal momentum jet model family reported in the aforementioned work. The dot-dashed line is given by the discrete model with liquid-only mass loss detailed in section \ref{sub:LOML} for the particular case of the ``submerged jet'' (air jet) as considered in the corresponding experiment. We can thus see that the LOML model gives the closes fit, increasing the precision, at the cost of losing the analytical solutions.

\section{Conclusions}\label{sec:conclusions}

We have presented 1D mathematical models of a conical jet applicable to the dynamics of a wide class of turbulent atomizing jets. The models' main assumptions include the so-called Locally Homogeneous Flow (LHF) for a two-phase flow. The models are based on conservation laws of the energy and the mass, and describe the dynamical quantities of the jet, viz. the density, velocity and dynamic pressure, along the jet's axis and are an extension and generalization of the ones found in \cite{FrancoMedrano-etal2017}. The main advantages of the proposed models over others in the literature are that the solutions are analytical explicit or implicit, they contain a single critical free parameter, viz. the jet's cone angle. We have previously proposed how to determine this angle so as to closely fit the experimental measurements \cite{FrancoMedrano-etal2017}. A couple of extra free parameters allow for fine tuning and have a physical interpretation leading to the possibility of being calculated indirectly from experimental measurements. More data is needed in order to test the accuracy and range of applicability of our theory. Special cases include a liquid jet in a very thin atmosphere and a submerged jet.


\section*{Acknowledgements}
FFM thanks the support of the Ministry of Education, Sports, Science and Technology of Japan, the Bank of Mexico's FIDERH program, the Mexican National Council of Science and Technology (CONACYT), Kyushu University's Graduate School of Mathematics and KUMIAY Internacional Co. Ltd.
This research has also been partially supported by the Center of Innovation Program from the Japan Science and Technology Agency, JST. YF was supported in part by a Grant-in-Aid for Scientific Research from the Japan Society of Promotion of Science (Grant No. 16K05476).

\bibliographystyle{plainnat}

\bibliography{jfm2017}

\begin{thebibliography}{31}
\providecommand{\natexlab}[1]{#1}
\providecommand{\url}[1]{\texttt{#1}}
\expandafter\ifx\csname urlstyle\endcsname\relax
  \providecommand{\doi}[1]{doi: #1}\else
  \providecommand{\doi}{doi: \begingroup \urlstyle{rm}\Url}\fi

\bibitem[Bidone(1829)]{Bidone}
George Bidone.
\newblock {Experi{\'e}nces sur la forme et sur la direction des veines et des
  courants de'eau lanc{\'e}s par diverses ouvertures}.
\newblock pages 1--136, 1829.

\bibitem[Birouk and Lekic(2009)]{Birouk-Lekic2009}
Madjid Birouk and Nebojsa Lekic.
\newblock {Liquid Jet Breakup in Quiescent Atomosphere: A Review}.
\newblock \emph{Atomization and Sprays}, 19\penalty0 (6):\penalty0 501--528,
  2009.

\bibitem[Cossali(2001)]{Cossali}
G.~E. Cossali.
\newblock {An integral model for gas entrainment into full cone sprays}.
\newblock \emph{J. Fluid Mech.}, 439:\penalty0 353--366, 2001.

\bibitem[Desantes et~al.(2005)Desantes, Payri, Salvador, and
  Gil]{Desantes-etal05}
J.~M. Desantes, R.~Payri, F.~J. Salvador, and A.~Gil.
\newblock {Development and validation of a theoretical model for diesel spray
  penetration}.
\newblock \emph{Fuel}, 85\penalty0 (2006):\penalty0 910--917, 2005.

\bibitem[Desantes et~al.(2011)Desantes, Salvador, L{\'o}pez, and {De La
  Morena}]{Desantes-etal11}
J.~M. Desantes, F.~J. Salvador, J.~J. L{\'o}pez, and J.~{De La Morena}.
\newblock {Study of mass and momentum transfer in diesel sprays based on X-ray
  mass distribution measurements and on a theoretical derivation}.
\newblock \emph{Exp. Fluids}, 2011\penalty0 (50):\penalty0 233--246, 2011.

\bibitem[Drazin and Reid(2004)]{Drazin}
P.~G. Drazin and W.~H. Reid.
\newblock \emph{{Hydrodynamic Stability}}.
\newblock Cambridge University Press, 2nd edition, 2004.

\bibitem[Faeth(1983)]{Faeth83}
G.~M. Faeth.
\newblock {Evaporation and Combustion of Sprays}.
\newblock \emph{Prog. Energy Combust. Sci.}, 9:\penalty0 1--76, 1983.

\bibitem[Faeth(1987)]{Faeth87}
G.~M. Faeth.
\newblock {Mixing, Transport and Combustion in Sprays}.
\newblock \emph{Prog. Energy Combust. Sci.}, 13:\penalty0 293--345, 1987.

\bibitem[Faeth et~al.(1995)Faeth, Hsiang, and Wu]{Faeth-etal95}
G.~M. Faeth, L.-P. Hsiang, and P.-K. Wu.
\newblock {Structure and Breakup Properties of Sprays}.
\newblock \emph{Int. J. Multiphase Flows}, 21:\penalty0 99--127, 1995.

\bibitem[{Franco} et~al.(2017){Franco}, Fukumoto, Velte, and
  Hod\v{z}i{\'c}]{FrancoMedrano-etal2017}
Fermin {Franco}, Yasuhide Fukumoto, Clara~Marika Velte, and Azur
  Hod\v{z}i{\'c}.
\newblock {Mass entrainment rate of an ideal momentum turbulent round jet}.
\newblock \emph{J. Phys. Soc. Jpn.}, 86\penalty0 (034401), mar 2017.

\bibitem[Fuchimoto et~al.(2009)Fuchimoto, Yanase, Mizushima, and
  Senda]{Fuchimoto-etal09}
Tetsuya Fuchimoto, Shinichiro Yanase, Jiro Mizushima, and Jiro Senda.
\newblock {Dynamics of vortex rings in the spray from a swirl injector}.
\newblock \emph{Fluid Dyn. Res.}, 41\penalty0 (4):\penalty0 045503, May 2009.

\bibitem[Ghosh and Hunt(1994)]{Ghosh-Hunt94}
S.~Ghosh and J.~C.~R. Hunt.
\newblock {Induced air velocity within droplet driven sprays}.
\newblock \emph{Proc. R. Soc. London, Ser. A}, 444\penalty0 (1920):\penalty0
  105--127, January 1994.

\bibitem[Gorokhovski and Herrmann(2008)]{Gorokhovski2008}
Mikhael Gorokhovski and Marcus Herrmann.
\newblock {Modeling Primary Atomization}.
\newblock \emph{Annu. Rev. Fluid Mech.}, 40\penalty0 (1):\penalty0 343--366,
  2008.
\newblock \doi{10.1146/annurev.fluid.40.111406.102200}.

\bibitem[Jiang et~al.(2010)Jiang, Siamas, Jagus, and
  Karayiannis]{Jiang-etal2010}
X.~Jiang, G.~A. Siamas, K.~Jagus, and T.~G. Karayiannis.
\newblock {Physical modelling and advanced simulations of gas-liquid two-phase
  jet flows in atomization and sprays}.
\newblock \emph{Prog. Energy Combust. Sci.}, 36:\penalty0 131--167, 2010.

\bibitem[Lin and Reitz(1998)]{Lin-Reitz1998}
S.~P. Lin and R.~D. Reitz.
\newblock {Drop and spray formation from a liquid jet}.
\newblock \emph{Annu. Rev. Fluid Mech.}, 30:\penalty0 85--105, 1998.

\bibitem[Liu(2000)]{Liu2000}
Huimin Liu.
\newblock \emph{{Science and Engineering of Droplets}}.
\newblock Noyes Publications / William Andrew Publishing, 2000.

\bibitem[McCarthy and Molloy(1974)]{McCarthy-Molloy1974}
M.~J. McCarthy and N.~A. Molloy.
\newblock {Review of Stability of Liquid Jets and the Influence of Nozzle
  Design}.
\newblock \emph{Chem Eng. J.}, 7:\penalty0 1--20, 1974.

\bibitem[Pastor et~al.(2008)Pastor, L{\'o}pez, Garc{\'i}a, and
  Pastor]{Pastor-etal08}
Jos{\'e}~V. Pastor, J.~Javier L{\'o}pez, Jos{\'e}~M. Garc{\'i}a, and
  Jos{\'e}~M. Pastor.
\newblock {A {1D} model for the description of mixing-controlled inert diesel
  sprays}.
\newblock \emph{Fuel}, 87\penalty0 (13):\penalty0 2871--2885, 2008.

\bibitem[Plateau(1873)]{Plateau}
J.~A.~F. Plateau.
\newblock {Statique exp{\'e}rimentale et th{\'e}orique des liquides soumis aux
  seules forces mol{\'e}culaires}.
\newblock \emph{Gauthier-Villars, Paris}, Vol. II:\penalty0 231, 1873.
\newblock Cited by Lord Rayleigh in Theory of Sound, Vol. II, Dover, New York,
  1945.

\bibitem[Rabadi et~al.(2007)Rabadi, Friedel, and Surma]{Rabadi-etal07}
Said~Al Rabadi, Lutz Friedel, and Robert Surma.
\newblock {Prediction of Droplet Velocities and Rain out in Horizontal
  Isothermal Free Jet Flows of Air and Viscous Liquid in Stagnant Ambient Air}.
\newblock \emph{Chem. Eng. Tech.}, 30\penalty0 (11):\penalty0 1546--1563, 2007.

\bibitem[Rayleigh(1878)]{Rayleigh1878}
J.~S.~W. Rayleigh.
\newblock {On the instability of jets}.
\newblock \emph{Proc. Lond. Math. Soc.}, 10:\penalty0 4, 1878.

\bibitem[Savart(1833)]{Savart}
F.~Savart.
\newblock \emph{Anal. Chem.}, 53:\penalty0 337, 1833.

\bibitem[Sazhin(2006)]{Sazhin2006}
S.~Sazhin.
\newblock {Multiple scales in spray modelling}.
\newblock \emph{J. Phys. Conf.}, 55:\penalty0 191--202, 2006.

\bibitem[Sazhin et~al.(2001)Sazhin, Feng, and Heikal]{Sazhin2001}
S.~S. Sazhin, G.~Feng, and M.~R. Heikal.
\newblock {A model for fuel spray penetration}.
\newblock \emph{Fuel}, 80:\penalty0 2171--2180, 2001.

\bibitem[Sazhin et~al.(2005)Sazhin, Crua, Hwang, No, and Heikal]{Sazhin2005}
Sergei Sazhin, Cyril Crua, Jin-Sik Hwang, Soo-Young No, and Morgan Heikal.
\newblock {Models of fuel spray penetration}.
\newblock In \emph{{Proc. Estonian Acad. Sci. Eng.}}, volume~11, pages
  154--160, 2005.

\bibitem[Stevenin et~al.(2016)Stevenin, Vallet, Tomas, Amielh, and
  Anselmet]{Stevenin2016}
C.~Stevenin, A.~Vallet, S.~Tomas, M.~Amielh, and F.~Anselmet.
\newblock {Eulerian atomization modeling of a pressure-atomized spray for
  sprinkler irrigation}.
\newblock \emph{Int. J. Heat and Fluid Flow}, 57:\penalty0 142--149, 2016.
\newblock ISSN 0142-727X.
\newblock \doi{10.1016/j.ijheatfluidflow.2015.11.010}.
\newblock URL
  \url{http://www.sciencedirect.com/science/article/pii/S0142727X15001502}.

\bibitem[Taylor(1962)]{Taylor1962}
G.~I. Taylor.
\newblock \emph{{Generation of ripples by wind blowing over viscous fluids}},
  volume~36.
\newblock Cambridge University Press, 1962.

\bibitem[Tomotika(1935)]{Tomotika1935}
S.~Tomotika.
\newblock {On the instability of a cylindrical thread of a viscous liquid
  surrounded by another viscous fluid}.
\newblock \emph{Proc. R. Soc. London, Ser. A}, 150:\penalty0 322--337, 1935.

\bibitem[Vallet et~al.(2001)Vallet, Burluka, and Borghi]{Vallet2001}
A.~Vallet, A.~A. Burluka, and R.~Borghi.
\newblock {Development of an Eulerian model for the atomization of a liquid
  jet}.
\newblock \emph{Atomization and Sprays}, 11\penalty0 (6), 2001.
\newblock ISSN 1044-5110.

\bibitem[Wakuri et~al.(1960)Wakuri, Fujii, Amitani, and Tsuneya]{Wakuri1960}
Yutaro Wakuri, Masaru Fujii, Tatsuo Amitani, and Reijiro Tsuneya.
\newblock {Studies on the penetration of fuel spray in a diesel engine}.
\newblock \emph{Bull. Japan Soc. Mech. Eng.}, 3\penalty0 (900475):\penalty0
  123--130, 1960.

\bibitem[Weber(1931)]{Weber1931}
C.~Z. Weber.
\newblock {Zum Zerfall eines Flussigkeitsstrahles}.
\newblock \emph{Math. Mech.}, 11:\penalty0 136--154, 1931.

\end{thebibliography}

\end{document}